# Systematic Review of Experimental Paradigms and Deep Neural Networks for Electroencephalography - Based Cognitive Workload Detection

Vishnu KN, *Student Member, IEEE*, Cota Navin Gupta, *Member, IEEE*

*Abstract*—This article summarizes a systematic review of the electroencephalography (EEG) - based cognitive workload (CWL) estimation. The focus of the article is two-fold, identify the disparate experimental paradigms used for reliably eliciting discreet and quantifiable levels of cognitive load and the specific nature and representational structure of the commonly used input formulations in deep neural networks (DNNs) used for signal classification. The analysis revealed a number of studies using EEG signals in its native representation of a two-dimensional matrix, for offline classification of CWL. However, only a few studies adopted an online/pseudo-online classification strategy for real–time CWL estimation. Further, only a couple of interpretable DNNs and a single generative model were employed for cognitive load detection till date during this review. More often than not, researchers were using DNNs as black-box type models. In conclusion, DNNs prove to be valuable tools for classifying EEG signals, primarily due to the substantial modeling power provided by the depth of their network architecture. It is further suggested that interpretable and explainable DNN models must be employed for cognitive workload estimation since existing methods are limited in the face of the non-stationary nature of the signal.

## I. INTRODUCTION

Brain–Computer Interfaces (BCIs) are often employed to facilitate Human – Machine Interactions (HMIs) like that of autonomous or semi–autonomous transportation vehicles or heavy industrial machinery. Operational aspects of these environments demand situational awareness, optimal allocation of attentional resources, and sustained vigilance from the operator due to their safety–critical nature [1]. These cognitive resource demands induce a load on the mental faculties of the human operator, and this operational load has been termed as cognitive (mental) workload (CWL). The cognitive resources demanded by an operational task may vary from very low (underload) at times to extremely high (overload) in ominous operational situations. High and low CWL may adversely affect the interaction, reducing both the machine's and operator's performance, which may result in catastrophes and cost human lives [2] Therefore, accurate real-time estimation of task-induced workload and the user's cognitive state is critical for an adaptive – automated functional protocol in real-world HMIs like piloting an aircraft, driving an automobile, or operating heavy construction machinery. Emerging technologies like Brain - Computer Interfaces (BCIs) are envisioned to bridge the gap between humans and machines by providing a bio-digital interface between the two [3], [4].

The general definition of CWL is the ratio between the cognitive resources demanded by the task and the available cognitive resources that a user can allocate against the task's demands [5]. Several ways of measuring task-induced CWL exist [6]. The field has traditionally used self-reported (subjective) measures to estimate the cognitive workload experienced by a user in addition to reaction time in a secondary task (a behavioral measure). These methods hinder the primary task execution and therefore are unsuitable for real-time estimation [7]. The adoption of neurophysiological signals such as electroencephalogram (EEG) has increased since they can provide an objective, direct, passive, and real-time estimation of the cognitive resources demanded by the task [8].

Electroencephalographic (EEG) signals originate from a noisy nonlinear system and have traditionally been considered challenging to decode [9]. Nevertheless, EEG is still an appropriate signal for CWL estimation [10] since it is a low-cost and portable acquisition system with high temporal resolution. However, this neuroimaging modality comes with a unique set of challenges. The high dimensionality of the EEG signal has always compelled feature extraction and dimensionality reduction [9], [11]from the time-domain signal, followed by dimensionality reduction. Other than physiological artifacts [12], the presence of unrelated neural activity could be the primary reason for EEG signals being highly variable across the multiple sessions of a subject and the different subjects performing the same task. Almost all state-of-the-art BCI protocols need extensive calibration for reliable classification performance at the levels typically required by consumer BCI applications [13]. These challenges necessitate careful experimental design and extensive signal processing before conducting statistical analysis, so it would be possible to correlate the EEG signal with an observed behavioral phenotype.

CWL can be elicited using numerous tasks and may be detected as changes in the signal power for various frequency bands of EEG signals. Many studies have independently verified characteristic changes in EEG sub-band oscillations during different levels of workload [14]. Alpha oscillations are characteristic of the wakeful state [15] since they relate to sensory perception and mediate sensory suppression mechanisms during selective attention [16]. Additionally, CWL can be measured using active or passive measures and tasks [17], [18]. A wide variety of these EEG-based measures have been reviewed in [19]. The passive BCI (pBCI) do not employ covarying subjective or behavioral measure of CWL. The envisioned pBCI is a bio-digital interface that can provide an implicit mode of communication between a computer-controlled machine through [8] by automatically detecting neurophysiological signals of specific intentions and translating this brain activity into machine-executable commands [20].

Identifying and segregating neural activity of interest from the rest of the signal is central to a BCI protocol, but the technical challenges

This work was submitted for review on 10th of June, 2023. This work was supported by the NEWGEN-IEDC fund from Department of Science and Technology, Govt. of India.

VKN was associated with the Indian Institute of Technology, Guwahati, Assam, India (kvishnu@iitg.ac.in). CNG was associated with the Indian Institute of Technology, Guwahati, Assam, India (cngupta@iitg.ac.in).

significantly hamper practical signal classification [21]. Recent advances in deep neural networks (DNNs) have shown promise in objectively assessing CWL levels from electroencephalogram signals [22], [23]. Deep learning (DL) algorithms can learn from the characteristically weak EEG signals eliminating the need for feature extraction [24] as well as signal pre-processing in some cases [25]. DNNs possess superior pattern recognition abilities over traditional machine learning algorithms (MLA) since they can leverage the parametric depth of the network while learning, enabling them to recognize the relevant features directly from the EEG signals despite the non-stationarity. In several EEG-Based BCI paradigms, (DNNs have surpassed the performance of traditional MLA [25]. Notwithstanding some sparse success in the field of CWL estimation [26], DNNs currently perform inferior to the current state-of-the-art SVM classifiers [22], [27]. However, it is worth noting that [28] proved that DNNs could achieve performance on par with traditional classifiers using relatively small EEG datasets.

Current limitations of EEG-based CWL detection and thus evident broad research gaps can be identified as,
1) Inter-session/subject variations in signal features notwithstanding the same stimulus being used for eliciting a given activity,
2) Non-stationarity of EEG signals and the need for signal features that deliver optimal classification performance for a given task,
3) Lack of models explaining the sustained inter-subject similarity in neural activity despite significant intra-subject signal variability [29], and
4) A consequent lack of consensus on the classification algorithm and the most appropriate signal feature.

Cognitive workload and measurements are currently widely used in aviation [30], automobile [31], and certain BCI applications. The field uses both laboratory and real-world-based paradigms. These experimental paradigms, DNN based detection methods, and their application domains are reviewed in this article. Though many reviews exist on the topic [2], [24], [30], [32], a systematic review focusing on EEG-based cognitive workload estimation using deep learning algorithms is absent, and this review is intended to fill this gap in knowledge.

## II. LITERATURE SEARCH

### A. Research Questions

The central topics of this review, the keywords for article retrieval, and The PRISMA flow chart of article selection are depicted in Fig. 1. We have identified 64 articles that satisfied all set criteria and have been analyzed using the following constructs. The research articles are systematically evaluated with predefined critical constructs expressed as questions
1) What are the paradigm designs used to elicit different cognitive states? Are there any domain-specific cognitive states and task-design trends?
2) What are the DNNs employed for cognitive load detection? What are the preffered network architecture, input formulations AND features used for CWL detection?

## III. RESULTS

### A. Experimental Paradigm for CWL Induction

Many experimental paradigms are prevalently used in CWL research to elicit graded levels of cognitive load. These levels vary from basic binary distinction to as high as seven levels of cognitive load. The highest levels of graded workload were provided by Automated Cabin Air Management System (AutoCAMS) task [33]. This experimental paradigm simulates the micro-world of a space flight but is a generic operator paradigm [34] where the subject is tasked to monitor gauge levels and make real-time decisions. The task's general similarity to many operational scenarios, including industrial process controllers, has inspired many studies, and about 10% of all the studies reviewed here are found using it. AutoCAMS simulates these graded levels by varying the number of subsystems and automation failures. The observed maximum in this survey is seven levels of cognition, as found in [35]. Most of the studies employing the AutoCAMS task used three or more categories. It is a computer simulation aimed at simulating adaptive operational environments. Additionally, different types of flight simulations were used to generate four [36]–[38] and five levels [39], [40] of workload. Further, the Multi-Attribute Task Battery (MATB ) [41] is also a generic operator paradigm like AutoCAMS, and it simulates the generic operations a pilot performs while flying an aircraft. The task battery consists of multiple tasks to be performed simultaneously in a scheduled manner, determining the induced workload levels. MATB has also been widely used for eliciting two [42] or three [28] workload levels.

Furthermore, Simultaneous Task Capacity (SIMKAP) [43]–[45] and N-back tasks [46] have been used to produce a maximum of three workload levels. SIMKAP task is a multitasking paradigm, and few open-source datasets collected with it are available online. Additionally, the N – back task consists of presenting a series of numbers or shapes to the subject on a screen, and the subject is asked to react to a series of stimuli by evaluating if the current element was the same as that appeared n times ago, and hence the name, n-back. Other paradigms used to elicit two levels of workload: working memory task, mental arithmetic task, construction activities, and learning tasks, to name a few. Apart from these standard tasks, some studies have used in-house tasks [47]–[49] for eliciting cognitive workload [50].

### 1) Cognitive and Operative Paradigms

A recent review classified the workload-inducing tasks into 'cognitive paradigms' and 'operate paradigms' [24] (Fig. 1)in which the studies using operative paradigms specifically intended their research as a direct industrial application, unlike the cognitive paradigm, which was intended as controlled laboratory experiments that are focused on theoretical aspects of cognition and the cognitive workload construct. This analysis follows the same taxonomical classification while presenting results. Most articles retrieved for this survey implemented operative paradigms for inducing cognitive workload (67%), while the rest were cognitive paradigms (33%). This inference is based on Fig. 2, where the pie charts illustrate the prevalence of the experimental paradigms used to elicit cognitive workload. The prevalent cognitive paradigms encountered in this study are N-back [51]–[54], Sternberg working memory [39], [55] mental arithmetic (MA) [56], [57], SIMKAP [58], [59], while General Flight Simulation [60]–[62], Driving Simulations [63]–[65], MATB [28], [42], [66] and AutoCAMS [67]–[72] are the prominent tasks categorized in operative paradigms. Overall, operative paradigms were encountered more in comparison to cognitive paradigms.

Following the logic of the previous synthesis, a further sub-division of these experimental paradigms is proposed keeping the objective of the cognitive load-inducing task and orientation of the conducting research. This classification of experimental task demarcates the orientation and application of the intended experiment, whether the construct under experiment is the human agent and his continually varying cognitive states or if it is an operational aspect of the machine that may bring about characteristic cognitive states in the user upon

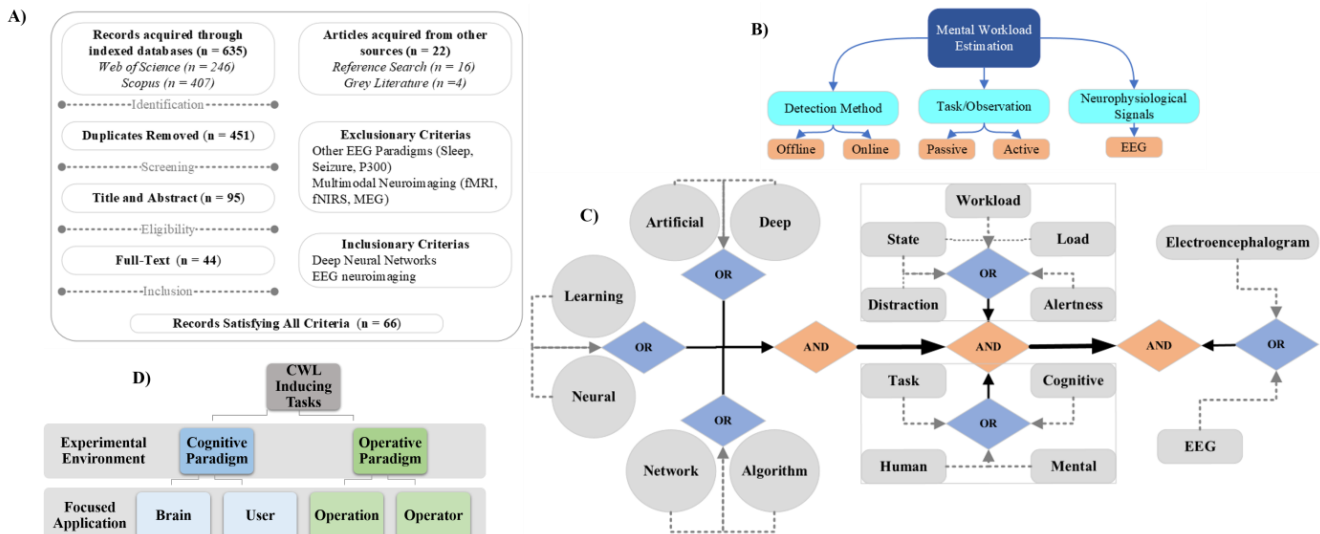

Fig. 1: A) The PRISMA Protocol followed in this review. B) The tree diagram of topics covered in this review. C) The set of keywords identified for each branch of the topic tree. The synonyms of a concept are joined using OR blocks and the different sub-themes are joined using AND blocks to construct the final search string D) The proposed taxonomical classification of CWL experimental paradigms

encounter. This classification of experimental tasks may help identify the specific context of the cognitive workload problem and help formulate it into a suitable experimental setting for the intended application. The taxonomical classification schemes are the 'operator paradigm,' 'operation paradigm,' 'user paradigm,' and the 'brain paradigm.' This dichotomy is depicted in Fig. 1. D. The results briefed in this section are depicted in Fig. 2

*2) Operation and Operator Paradigms*

The operator paradigms simulate the general characteristics of an HMI, focusing on the operator, while the operation paradigm focuses on simulating one particular aspect of a given HMI to examine the corresponding functional states of the human agent. About 47% of all the articles surveyed use an operator paradigm, while about 20% of the studies employed an operation paradigm. Within the operator paradigms, flight simulations that mimicked the typical operational environment of a pilot were used the most (30%). Monotonous automobile driving (19%) and the generic operator paradigm AutoCAMS (19%) were the next most prevalent operator paradigms. Further, MATB was used by 16% of the studies. Other operator paradigms are driving in varying traffic conditions [73], construction activities [74], [75], and learning tasks [76], and together they constituted about 16% of the operator paradigms encountered in this survey. Contrary to the operator paradigms, operation paradigms were focused on inducing a specific cognitive state in response to a particular operational sequence or event, such as lane - deviation. The most prevalent experimental task within operation paradigms was the lane-deviation task (46%), where a lane-perturbation event was followed by monotonous automobile driving, and the operator's reaction time was regressed against the driver's cognitive state. Other operation paradigms encountered in this survey are driving distraction [77], remote piloting of aerial vehicles [78], specific flight sequences [36]–[38], robot-assisted surgery [79], and construction activity [80], constituting about half of the operation paradigms (54%).

*3) Brain and User Paradigms*

User paradigms focus on user skills or specific attributes of the user, like multitasking or language proficiency, while brain paradigms focus on cognition-related aspects such as working memory (WM) and engagement. About 18% of all the reviewed articles induce workload with a user paradigm, while brain paradigms were used by 15% of all the articles. The prevalent user paradigm was MA (46%), where the subject continuously performs difficult ssssnumerical calculations to elicit binary workload levels, followed by SIMKAP (38%), where several sub-tasks are performed simultaneously to elicit graded workload levels. Other user paradigms include visuomotor tracking tasks [81], where a visual stimulus is tracked while moved through a screen, and language incongruency tasks, where ambiguous pronouns elicit higher workload levels. These tasks focused on the user and their response to a generic BCI protocol.

Further, within the brain paradigms, the N-back task (46%) was the prevalent choice, the rest (56%) were several types of WM tasks and other in-house WM paradigms. These tasks were focused on a specific aspect of cognition, such as WM, attention, arousal, and vigilance.

*4) Experimental Environment*

Usually, computer-based simulations were used to set up task environments in cognitive and operative paradigms. MATB and AutoCAMS are two computer–based operator paradigms extensively used for eliciting cognitive load, and they resemble the typical operational environment of an aircraft pilot and a generic industrial process controller, respectively. Unlike computer-based simulations, some deep-learning studies used EEG signals acquired from real-world vehicle operations scenarios [50], [82]. However, simulated task conditions are the norm in the field. Typically, these tasks are implemented in an augmented or virtual reality engine or a computer-based simulator. Though cognitive paradigms were created using only computer-based simulations, task environments of operative paradigms were set up much more diversely.

In the automobile industry, augmented reality (AR) 'full–driving simulators' are less of an industry standard and are typically used only in research settings. The full-driving simulators constitute a real car mounted on a Stewart platform that can simulate motion with six degrees of freedom and surrounding projected display. AR environments are industry standard for pilot training in the aviation sector. These systems are known as 'full–flight simulators' and vary in terms of the realism induced in the simulation they offer. On the high end, data collected using full-flight simulators were encountered in this study [61], [83]. On the lower end, this study used a simple simulation setup like mounting the pilot's chair on a Stewart platform [39], [84] and providing computer– based projected display. Additionally, virtual reality (VR) engines and head-mounted displays were used only for conducting construction activity paradigms[74].

These AR and VR systems are a good trade-off between real-life situations and controlled laboratory environments and can be extremely useful in researching real-life paradigms which are often dangerous, like that of a lane-keeping task. However, unlike these costly systems that are not easily available, computer–based simulations are accessible to everyone. Moreover, many studies have validated computer–based operator paradigms like AutoCAMS and MATB, and there already exists a plethora of datasets collected using these tasks, and therefore it can be used for comparing the fidelity of detection methodologies.

*B. Cognitive States Induced by CWL*

The cognitive state of arousal, characterized by attention and engagement, is achieved when workload levels are optimized. Cognitive States and varying degrees of workload levels were seen across all the studies reviewed here. It was enquired whether any experimental paradigm was preferentially used to elicit a given cognitive state. It was observed that specific cognitive states were induced by domain-specific experimental paradigms. AutoCAMS and MATB were particularly suited for generating highly graded CWL levels due to their highly modular nature. Notably, different workload levels were elicited by 47% of the studies reviewed here. The states of attention and engagement have been explored by 10% of the studies. Overload fatigue was examined by about 16% of the studies, while underload fatigue was explored by 18%. Further, WM was explored by about 9% of the studies. These results are described in Fig 2. G

Moreover, underload fatigue was mostly explored in automobile paradigms since detecting drowsy states is a popular domain-specific industrial need. On the other hand, operational fatigue was mostly explored in aviation paradigms. Apart from specific flight sequence simulations, only AutoCAMS was used to elicit operational exhaustion and overload fatigue. It is interesting to note that only brain paradigms explored WM, operation paradigms explored underload fatigue (drowsiness), and operator paradigms induced operational fatigue, while all types of paradigms explored attention, engagement, and multitasking abilities.

*C. Deep Neural Networks for CWL Detection*

There were mainly two kinds of studies that used a DNN for CWL detection, those that treated the model as a black box [28], [79] and those who have reasoned out the architecture and pipeline [58].. However, other studies have modified parts of the architecture and pipeline to suit the specific problem of EEG – based CWL detection [45], [51]. Most networks were implemented offline, and only two studies were found to use online pipeline [56], [81] explicitly. However, other studies [56], [74], [85] employed a pseudo – online analysis. Further, one study was found implementing a CWL detection system on a smartphone.

Some studies (about 25%) have introduced additional DL mechanisms like attention [86], residual identity connections [54], or multi-paths [42], [45] to endow the network with additional modeling power. Within the group of studies employing additional DL mechanisms, residual connections, commonly known as ResNets were the prevalent (29%) choice [75], [87], followed by the attention mechanism [45], [54], [86] with about 17% prevalence. ResNets are are generally used to solve the vanishing gradient problem. Attention, on the other hand, enables the network to focus only on the parts of the input relevant to the problem at hand and is a method known to improve computational burden and performance. Attention has been used for feature selection in some studies, where it is employed before the network input layer, but most employed this mechanism within their network In the latter case, the output of several deep learning layers containing high-dimensional features are transformed with weighted multiplication of for determining their contribution to the each prediction.Further, about 17% of the studies reviewed here used both Ensemble Learning [70], [71] and Transfer Learning [69], [87]. Ensemble Learning trains several classifiers on sub-sets of the data and aggregates the information from all for making a prediction. The method is advantageous in the case of EEG since the signals are highly variable across sessions and subjects. One such study used an ensemble of AE networks to mitigate the cross-subject variability of EEG signals. It was enquired if any preference exists for DNNs in different application domains. It was observed that the most networks have been used in all types of experimental paradigms, and no such preference exists.

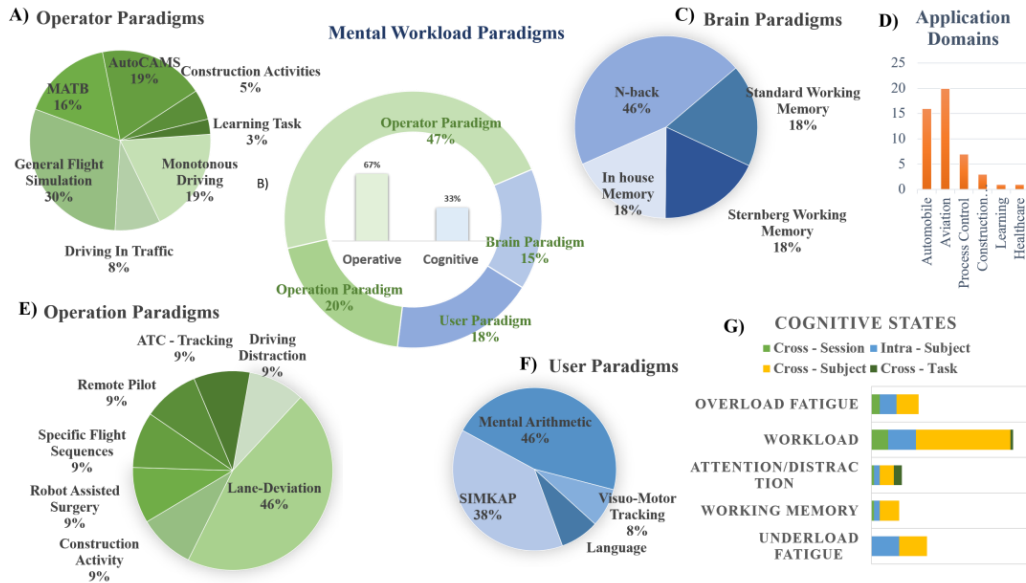

Fig. 2: The pie chart (B) describes the percentages articles using each task for collecting data. The pie charts have been organized to mimic the taxonomical bifurcation of cognitive and operative paradigm with color coded categories. The sub-charts show the distribution of operator (A) / user paradigms (F), and operation (E) / brain paradigms (C). The sizes of slices signify the prevalence within paradigm category. D) Depicts the application domains of CWL research. G) The bar charts depict the prevalence of different cognitive states encountered in this study, they are color coded to reflect the generalizability of the study, whether it is a cross-session/subject/task analysis.

*1) Network Architecture*

Convolutional Neural Networks (CNNs) are the most prevalent network of choice (29%), presumably due to their success in computer vision. Plug-and-play type architectures and availability has been credited in at least one of the studies for the motivation of using a CNN for cognitive workload detection [79]. The generalizability of CNNs in recognizing spatial patterns from data structured in a 2D / 3D matrix might have been a reason for the choice [88]. Recurrent Neural Networks (RNNs) were the next prevalent architecture (24%), and they were explicitly motivated by the recurrent nature of the network and its known capabilities of modeling temporal dependencies [28], [42], [54].

Hybrid Neural Networks (Hybrids) and Auto-Encoders (AE) were used by about 15% and 12% of the studies reviewed in the survey. The hybrid type of networks only used CNN – RNN combinations, and hybrid networks consisting of other networks and algorithms are not to be found in this systematic survey. Other architectures encountered in this survey are Multi-Layer Perceptron/ Artificial Neural Networks (MLP/ANN) [39], [63] (9%), Deep Belief Networks (DBNs) [68], [89] (8%), Generative Adversarial Networks [90] (GANs) (2%), and Graph Neural Networks [73] (GNNs) (2%). The prevalence of neural networks is given in Fig. 3 A.

*2) Signal Feature Extraction*

Popular features used in cognitive workload research can be categorized into five groups: '*spectral*,' '*nonlinear*,' '*temporal*,' '*spatial*,' '*statistical*,' and '*others*.' Most studies used a combination of features from these groups, and only a few chose only a single type of feature. Since studies used a combination of features, each article was counted separately against each feature. About 72% of the studies reviewed here used a feature extraction step before modeling EEG with a DNN, but about 23% of studies eliminated the feature extraction step and directly fed the EEG signals to the DNN for analysis. However, within the studies that used no specific feature extraction step, most employed some signal filter or artifact reduction methods to clean the signal for analysis, and very few studies directly used the raw EEG signal as input to the DNN [65].

Within the studies that employed feature extraction steps, about 54% of studies extracted various spectral features from the EEG signals. It usually included calculating power spectral density using various methods, including Fourier and discreet wavelet transforms. Specifically, the frequency bands of theta, alpha, and beta were extracted by most studies since they were known to be the most relevant channels for CWL detection [14]. Some studies used all frequency sub-bands; however, most studies eliminate gamma at the pre-processing stage by applying a high-pass filter that excludes gamma oscillations.

Nonlinear features, such as various entropy-based measures, were the next most prevalent feature (15%). These networks were motivated by the nonlinear behavior of EEG [29] and expected entropy measures to contribute to the classification performance significantly, and [28] found that their RNN performed slightly worse when nonlinear features were not given to the network. Most notably, approximate entropy [28], Shannon entropy [68], [69], Spectral entropy [69], nonlinear functional connectivity features [52], and mutual information [91] were used by the articles reviewed in this analysis. Generally, nonlinear features were fused or concatenated with other feature types before feeding to the network as in [28], while [52], [91] were found to training their classifier exclusively on nonlinear features. Further, some studies (11%) used statistical measures of the EEG signal like mean, variation, and kurtosis for training their networks. All statistical features were extracted by [76]; however, most studies that used statistical features for training their model typically extracted mean, variance, skewness, and kurtosis. Additionally, most studies concatenate all statistical features of interest before feeding them as a final input for the network.

About 10% of studies used temporal features other than the time-domain signal, such as the auto-regressive coefficients [89] and moving average algorithms. All temporally varying features except the time-varying frequency (spectral feature) and time domain signals (no feature) are grouped into this category. Other features combined was found in about 8% of the studies. Among them, one study explored the use of fractals [63], while two studies explored both functional connectivity [52] and graph features. A recent review exhaustively enlists all the popular EEG features extraction typically used in signal classification [11].

It was further enquired whether any feature was preferentially employed for a given DNN to see if the network architecture necessitates the feature extraction in addition to the complexity of the signal. It was observed that spectral features were used for all DNN architectures along with nonlinear measures, and since these features have strong theoretical foundations in EEG analysis, it is unsurprising that all networks used these two as input features. However, Graph Neural Networks (GNNs) and Generative Adversarial Networks (GANs) have not been found using these two measures, possibly due to their architecture. It was also observed that DBNs and AEs did not use temporal features or the time-domain signal, as they exclusively preferred a concatenated feature vector. This analysis is depicted in Fig. 3 A & B.

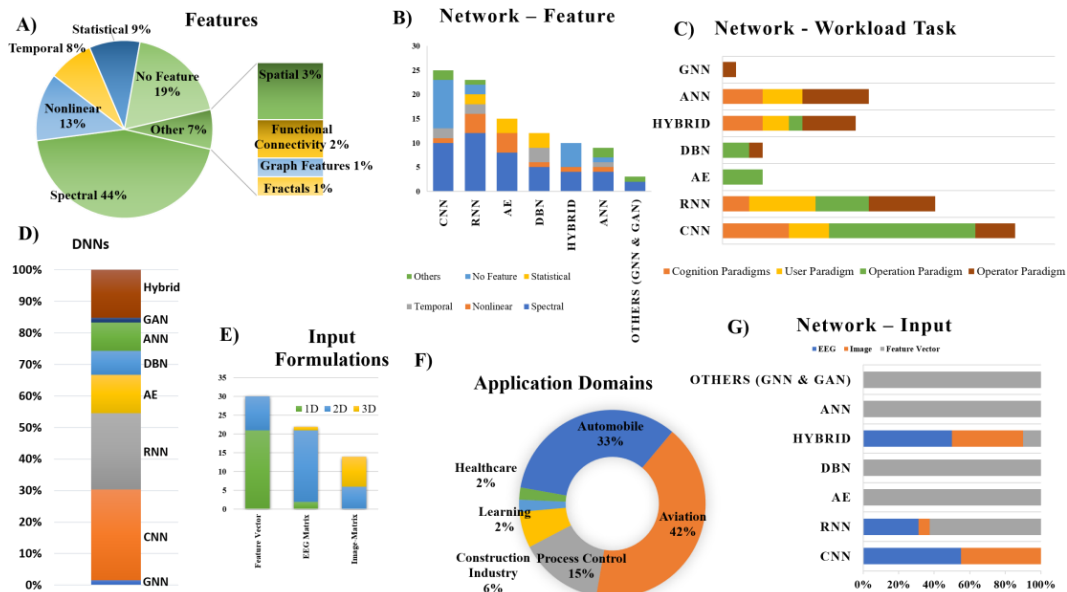

Fig. 3: D) The chart depicts the prevalence of DNNs encountered in this survey A) The plot describes the percentage of different feature used by the studies reviewed here. E) This bar chart depicts the input formulations used by the networks. 1D feature vectors were clearly the preferred input C) Chart depicts the paradigms specific choice of networks, according to proposed taxonomical grouping. It is clearly seen that there is no preference of network for any of the tasks. G) The preference of a network for a given input formulation is plotted. B) the preference of features among networks is plotted in this bar chart.

*3) Network Input Formulation*

There were three main categories of input formulations to be seen in this analysis, feature vector, image matrix, and EEG matrix. The feature vector is usually a concatenation of all the features in a suitable format for the employed DNN, while the image matrix is a single-channel or multi-channel image-like data created from the EEG signals using various signal transformation methods. The EEG matrix contained multi-channel signals in its native 2-dimensional (2D) form. Most studies that used multiple categories of features concatenated these into a suitable feature vector.  Overall, feature vectors were the most used input formulation, followed by image matrices and EEG matrices.  AEs, DBNs, and ANNs have all used exclusively feature vectors as input, while CNNs have not been trained using only feature vectors as input. This result is depicted Fig. 3 E.

Within the feature vector, there were 1-dimensional (1D) and 2D feature vectors, where the 1D feature vector is usually a concatenation of all the features extracted from the data without any specific sequential relationship amongst its elements like the ones used in [28], [67]. Only Spectral and nonlinear features were concatenated to create vectors in [28], [69], [92] to create 1D or 2D vectors, while statistical features were concatenated in addition to the above in [50], [68]. Additionally, feature vectors were created using spectral power density in different bands, a CWL index known as the fatigue index [37]. The 2D feature vectors are also a concatenation of features except when the whole spectral decomposition matrix was used.

In the image category, single-channel images (2D – images) [55], [61], [80], and multi-channel images (3D or higher) [75], [90], [93] were used as input to the network within the image category. These images were mostly created by transforming the time-domain EEG signal into the spectral domain. Many variations of image-like data were created from EEG signals using some topographical projection and interpolation for transforming the EEG data into a multi-channel or single-channel image. These methods mainly differed in the feature extraction step and the transformation used for mapping. [94], [95]. Certain studies employ spectral density at each location within a given time to produce a series of images, like a 'brain power map' [37]. Some studies have created an image-like representation by concatenating the spectral decomposition matrix of different frequency bands into the multiple channels of an image [53], thereby suggesting EEG-image is a general feature that may be used for EEG signal analysis.

EEG matrix was directly given as input for a given model under the assumption that DNNs can leverage their depth to model the inherently noisy EEG signals. Time-domain input was given mostly to RNNs and occasionally to CNNs.  Some studies have used EEG signals assuming them to be 2D images [79], [85]. This, however, is not entirely supported by the assumptions of the CNN models employed since the arrangement of channels (the spatial location of the signals) does not follow any reasonable pattern resembling their spatial location in the scalp. Some studies used a 1-dimensional (1D) EEG vector, while others used a concatenation of multiple 2D frames into a 3-dimensional (3D) EEG matrix. Some CNNs can perform depth-wise, channel-wise, separable, and/or dilated convolutions and are adapted to process temporal dependencies.

It was further enquired whether the DNNs favor any specific type of input formulations and whether there was any consideration to be given while creating the input vector for a given network. Time-domain inputs were exclusively used by MLPs, CNNs, and RNNs and their hybrids. Time-domain input is appropriate for RNN; for others, the insensitivity to temporal dependencies may prevent the time-domain signal from being useful to the network.. It was also observed that Image-like representation was only used for CNNs and their hybrids. These results are summarized in Fig 3. G

*4) Generalizability of the Network*

The least generalizable model is the subject-specific model that has been explored by 27% of the studies, and many of these studies have recorded EEG from only one session per subject.  Cross-Session models mark the next level of generalizability, and about 10% of studies have explored such a detection strategy. About 58% of the studies reviewed have proposed cross-subject classifiers, which suggest a high level of generalizability across the subjects and different sessions. However, it is notable that most studies have pooled multiple subjects/sessions with simplistic assumptions for training the data and have not considered the nonlinear statistical shifts present in the EEG signals from multiple sessions and subjects. The generalizability of cognitive states and the DNN detector is depicted in Fig. 3. G.

The highest level of generalizability is achieved by around 4% of the studies, as they have built models to recognize workload levels across different tasks [48], [54]. These classifiers may have accurately estimated universal discriminatory features of different cognitive workload levels. However, it is still unclear what the significant contributing factors for the predictions and decisions made by these networks, very few studies  [58], [96] have interpreted the networks' latent representations and attempted at explainable or interpretable deep learning.

## IV. DISCUSSION

The principal motivation for this systematic literature analysis was to identify the most suitable methods for elicitation (experimental paradigms) and detection (EEG-based DNNs) specific to the different application domains of CWL research. This analysis found no specific trends in the architectural choice or training strategy according to the tasks or the targeted cognitive states as expected. However, clear patterns were present regarding the types of features and the data structure used for training a DNN, as described in the results section.

Deliberations on the limitations of DNN-based detection lead to generalizability as overfitting is an imminent concern for any DNN; the peculiarities of EEG data only aggravate the issue. Some studies have built subject-specific classifiers since EEG is known to be having nonlinear statistical shifts across different subjects. These can be considered as the least generalizable models. Additionally, since EEG is a non-stationary signal across multiple sessions of a single subject, cross–session detection of cognitive workload is a challenging problem, and it might be caused by the fact that the number of recording sessions seen in typical EEG datasets might not offer enough modeling power for the network to capture variations across sessions. Most deep learning pipelines use a cross-subject training strategy to train the network. This trend may be attributed to the typical low sample sizes of EEG data, which would not offer sufficient samples from a subject to train a DNN since most studies did employ any mathematical transformation to bring the signals from multiple subjects into a shared dataspace and have pooled them indiscriminately. Therefore, it can be suggested that existing DNNs can already perform cross-subject classification, suggesting they offer sufficient generalizability to model users' cognitive workload levels. Further, some studies have attempted cross-task classification of workload levels using the same DNN [54], [97], [98]. The performance of these networks suggests that cognitive workload levels elicited by different tasks may elicit similar neural responses and that they can be detected using a deep neural network. In summary, DNNs offer sufficient generalizability for employing them for CWL-level detection across subjects and tasks, provided they are trained with sufficiently heterogeneous data.

A key issue identified in this survey is that of an appropriate input formulation. CNNs are particularly good at learning spatial relationships in a 2D matrix representation of data. However, since the EEG channels (matrix rows) are not arranged according to its (EEG electrodes) spatial location on the scalp, the EEG matrix does not adequately represent the spatial relationship between the channels. CNNs assume that the if the input data is one with spatial dependencies. Thus, the DNN cannot capture scalp spatial information when the native EEG matrix is presented to the network. Therefore, further measures of experimental controls need to be defined for employing CNNs directly on EEG matrices. A suggestion is to randomly change the location of EEG channels in the matrix representation and cross-validate the model. Further consideration of the input formulations for RNNs suggests that feeding a concatenated feature vector to an RNN is problematic since RNNs assume that the input vector's elements share temporal dependencies. Therefore, concatenating temporally unrelated features into a feature vector is unjustifiable in the case of an RNN. This issue has been correctly identified by [28], though they have used a set of temporally uncorrelated spectral and nonlinear features concatenated into a 1D vector.

Another key issue identified is related to the subjects of CWL experiments. In laboratory paradigms, the subjects are mostly graduate students. Aviation and automobile paradigm still possessed larger variability due to professionals being used as subjects. However, all other paradigms predominantly use university students, presumably because of availability. In most cases, subjects tend to be in the below-40 age bracket. However, cognitive workload is known to change with age. Therefore, one of the suggestions this review put forth is to include older and younger individuals alike in the subject pool.

Though many have deliberated on core problems of EEG-based cognitive state detections, solutions to these fundamental problems are still at large. This article postulates that deep neural network offers promising solutions to the challenges of EEG-based cognitive workload detection, such as automatic feature extraction and signal variability. Further, it is hypothesized that DNNs (using transfer learning) can overcome the domain statistical shifts in the EEG data across different sessions and subjects without using sophisticated data-pooling techniques [28], given that the training set is sufficiently large and heterogenous. Further, there were few online classifiers that may be useful in a practical BCI, though some have validated their framework using pseudo-online DNN designs. There was only one study that implemented their CWL framework in a smartphone. These findings suggest that a real-time framework needs extensive research to see if DNNs are a viable computing solution for real-time cognitive state detection in online BCI protocols.

## A. Cognitive Load Continuum

The bibliometric data presented in this article suggests there are two central themes dealt in the studies reviewed here, the overload or the underload of cognitive resources, and respectively the resulting drowsiness or distraction. Further, this systematic review theorizes a proposition termed 'the cognitive load continuum' where all the disparate cognitive states and associated workload levels are expressed as a function of cognitive workload demand and available cognitive resources for allocation using the existing multiple-resource theory [5]. Transient neurophysiological changes that lead up to a certain cognitive state, like that of fatigue, can be modeled as the state-transitory causes and effects in this framework. The proposition is graphically described in Fig. 4. A.

In an operational context, these operator functional states (OFS) can vary continually due to task-related affairs. The optimal OFS is hypothesized to be an unstable equilibrium in its cognitive landscape. Furthermore, sub-optimal OFS can result from being under cognitive load for a prolonged period, which can be termed fatigue. There are two types of fatigue. Fatigue from cognitive overload, or operational exhaustion, leads to sub-optimal operator performance as the attention resources are depleted due to physiological fatigue. Fatigue from cognitive underload, or drowsiness, also leads to sub-optimal operator performance as the attention resources are reduced due to mind wandering or preoccupancy to sleep. This relationship is depicted in Fig. 4. B.

## V. CONCLUSION

The general operator paradigm was simulated using AutoCAMS and MATB with highly graded workload levels. Further, it has been observed that specific paradigms were used for eliciting some cognitive states, though a wide variety of tasks were used for eliciting graded/binary workload levels. Notably, drowsiness and underload fatigue were explored more by automobile driving tasks, while operational exhaustion and overload fatigue was explored more often

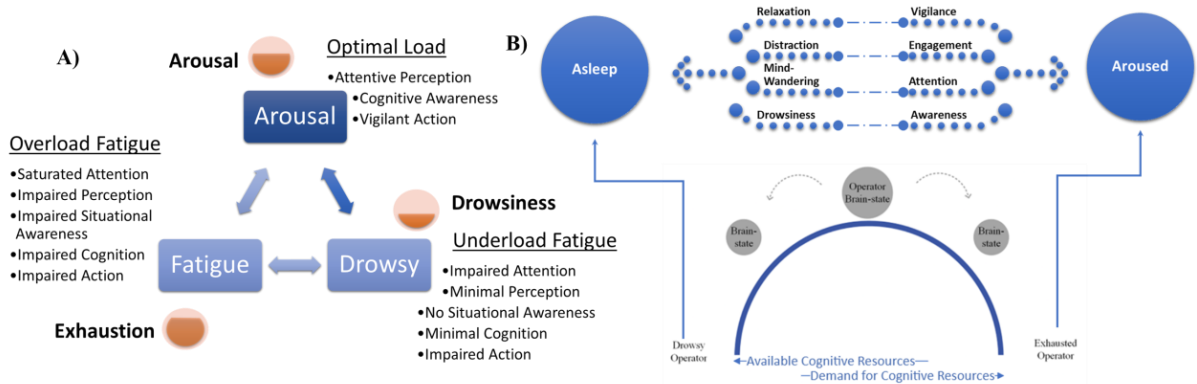

Fig. 4: A) The extreme ends of this scale are the extremes of cognition, the lower end being an unconscious state without any perception, cognition, or action. B) The curve depicted is the cognitive demand – and allocation curve on which the operator brain state achieves an unstable equilibrium of optimal cognitive load and delivers maximum performance. While on either side of this demand-allocation curve, the operator performance decreases

using aircraft flying tasks. Further, this analysis observed that broad-spectrum DNNs were used to detect all states in the cognitive load continuum. Moreover, there was no clear preference among the task choices or paradigms for DNNs. However, the input formulations and features were considerably biased toward some networks. AE and DBNs used all features equally and mostly used concatenated features, while CNNs preferred 2D EEG matrix or artificial images constructed out of it.

RNNs preferred time-domain input and did not quite nearly were not successful when other features were used. The preference is presumably explained by the temporal modeling abilities of RNNs. Transformer networks, another popular time-series modeling architecture was not to be found during this analysis. In the case of hybrid networks, mostly EEG time series and EEG-Images were used, and these networks were implemented to capture both the temporal and spatial dependencies in the multi-channel – EEG signal. Despite their underperformance, this review argues in favor of DNNs considering their generalizability and modeling power. Further, with the emergence of explainable DNN architectures and interpretation techniques that can be used to infer the features learned from the hidden layers of a DNN, theoretical studies may leverage them for modeling the phenomenology of neural activity.